\documentclass[12pt]{JHEP3}
\usepackage{epsfig}
\epsfclipon
\usepackage{multicol}
\usepackage{epsfig,bm}
\usepackage{amssymb,amsmath}
\usepackage{graphicx}

\newcommand{\be}{\begin{equation}}
\newcommand{\ee}{\end{equation}}
\newcommand{\bea}{\begin{eqnarray}}
\newcommand{\beas}{\begin{eqnarray*}}
\newcommand{\eea}{\end{eqnarray}}
\newcommand{\eeas}{\end{eqnarray*}}
\newcommand{\ba}{\begin{array}}
\newcommand{\ea}{\end{array}}
\def\ls{\mathrel{\lower4pt\vbox{\lineskip=0pt\baselineskip=0pt
           \hbox{$<$}\hbox{$\sim$}}}}
\def\gs{\mathrel{\lower4pt\vbox{\lineskip=0pt\baselineskip=0pt
           \hbox{$>$}\hbox{$\sim$}}}}

\newcommand{\vp}{\varphi}

\def\smiley{\hbox{\large$\bigcirc$\hspace{-.80em}%
\raise.2ex\hbox{$\cdot\cdot$}\kern-.61em    
\lower.2ex\hbox{\scriptsize$\smile$}}\ }

\newcommand{\roughly}[1]{\mathrel{\raise.3ex\hbox{$#1$\kern-0.85em
\lower1ex\hbox{$\sim$}}}}

\newcommand{\lsim}{\roughly<}

\def\be{\begin{equation}}
\def\beq\begin{equation}
\def\ee{\end{equation}}
\def\bea{\begin{eqnarray}}
\def\eea{\end{eqnarray}}

\def\beq{\begin{equation}}
\def\eeq{\end{equation}}
\def\beqa{\begin{eqnarray}}
\def\eeqa{\end{eqnarray}}

\newcommand{\bmat}{\left(\begin{array}}
\newcommand{\emat}{\end{array}\right)}

\title{Non-Gaussianity from Instant and Tachyonic Preheating}

\author{Kari Enqvist,$^{1,2}$ Asko Jokinen,$^{3}$ Anupam Mazumdar,$^{3}$
Tuomas Multam\"aki,$^{3}$ Antti V\"aihk\"onen,$^{1,2}$\\
 ${}^1$ Helsinki Institute of Physics, P.O. Box 64,
FIN-00014 University of Helsinki, Finland\\
 ${}^2$ Department of Physical Sciences,
P.O. Box 64, FIN-00014 University of Helsinki, Finland\\
 ${}^3$ NORDITA, Blegdamsvej-17, Copenhagen, DK-2100, Denmark}

\abstract{We study non-Gaussianity in two distinct models of
preheating: instant and tachyonic. In instant preheating
non-Gaussianity is sourced by the local terms generated through the
coupled perturbations of the two scalar fields. We find that the
non-Gaussianity parameter is given by $f_{NL}^{\phi}\sim 2g< {\cal
O}(1)$, where $g$ is a coupling constant, so that instant preheating
is unlikely to be constrained by WMAP or Planck. In the case of
tachyonic preheating non-Gaussianity arises solely from the
instability of the tachyon matter and is found to be large.  We find
that for single field inflation the present WMAP data implies a bound
$V_{0}^{1/4}/M_{P}\leq 10^{-4}$ on the scale of tachyonic instability.
We argue that the tachyonic preheating limits are useful also for
string-motivated inflationary models.}

\preprint{NORDITA-2005-03\\ HIP-2005-01/TH}

\begin{document}


\section{Introduction}

Preheating, first realized in~\cite{Robert} and worked out in
detail in Refs.~\cite{Many1,Linde,Dan}, is an interesting
possibility for obtaining a thermal Universe after the end of
inflation.  Preheating is a non-perturbative phenomenon and
perhaps the most efficient way of transferring the inflaton energy
density into other degrees of freedom. Without preheating the
inflaton would have to decay perturbatively and slowly, thereby
requiring a long time scale for the thermalization of the inflaton
decay products (for recent discussions, see~\cite{Jaikumar}).

The simplest realization of preheating is obtained if the inflaton
condensate has a coupling $g^2\varphi^2\sigma^2$, where $\varphi$
is the inflaton and $\sigma$ is another scalar field. Then during
the coherent oscillations of $\varphi$ a resonant
production of $\sigma$ quanta can take place due to a temporary
vacuum instability. The occupation number of $\sigma$ field
increases gradually while the fluctuations in $\sigma$ increase
exponentially. An important observation is that during this
phase gravitational (metric) fluctuations also get amplified at
super-Hubble scales~\cite{Many2,Easther} (for a general treatment of first
order linear perturbation theory, see~{\cite{MFB})
         
The growth of metric fluctuations is of great interest because of the
potential observational consequences. In particular, during inflation
the growth of the second order metric fluctuations leads to an
enhancement of the non-Gaussianity parameter $f_{NL}^{\phi}$ of the
CMB (for a review of non-Gaussianity, see \cite{ngrev}). In our recent
paper~\cite{we} we considered a particular preheating scenario where
in comparison with the inflaton, the dynamics of $\sigma$ was assumed
to have negligible impact on metric fluctuations. We estimated the
second order metric fluctuations following \cite{Acquaviva, Antti} and
found that $f_{NL}^{\phi}$ grows exponentially with a rate that
depends on the number of inflaton oscillations. (This is a generic
result and applies also to the case of supersymmetric flat directions,
which could provide an alternative scenario for reheating the
Universe~\cite{Enqvist}). The result suggests that non-Gaussianity may
provide an important test of preheating scenario, and in the present
paper we shall demonstrate that this is indeed so.

Non-Gaussianities can also arise after inflation~\cite{Riotto} when
the energy in the inflaton condensate is transferred to other degrees
of freedom.  Such a situation arises during preheating as we shall now
argue. Let us focus on two different cases. First, let us note that
the inflaton could decay non-perturbatively during just one
oscillation only. This is simple to understand for example in a
chaotic type model with $V\sim m^2\varphi^2/2$. There inflation ends
at $\varphi\sim M_{P}$ and the field rolls down towards its global
minimum. However, by virtue of the coupling, $g^2\varphi^2\sigma^2$,
the $\sigma$ field obtains an effective mass
$m_{\sigma}^2=g^2\varphi^2$. Initially, as long as $|\dot
m_{\sigma}|\ll m_{\sigma}^2$, the fluctuations in $\sigma$ evolve
smoothly. Eventually the adiabatic condition is violated, and particle
production occurs when $|\dot m_{\sigma}|\sim g\dot\varphi\geq
g^2\varphi^2$.  During the first oscillation the Hubble expansion rate
can be neglected, i.e. $m_{\varphi}>H(t)$. Hence the field velocity
near the minimum of the potential is given by $|\dot\varphi|\approx
m_{\varphi}\varphi_{e}$, where $\varphi_{e}\sim 0.1M_{P}$ is the
amplitude of the first oscillation after the end of inflation. In this
situation non-adiabatic condition is violated when
$-\varphi_{\ast}\leq \varphi\leq \varphi_{\ast}$, where
$\varphi_{\ast}\sim\sqrt{m_{\varphi}\varphi_{e}/g}$~\cite{Linde,Lindeinst}.
For a sufficiently large coupling, $g\gg 10^{-4}$, which is required
for an efficient reheating, the interval of non-adiabatic violation is
very short. This situation is known as {\it instant preheating} and
has many interesting cosmological consequences (for a discussion,
see~\cite{Lindeinst}).

The second case is obtained in models where the scalar field
undergoes instability due to the appearance of a tachyonic mass term
$-m^2\chi^2/2$ in the effective potential~\cite{Lindetach} which thus
induces symmetry breaking, where we assume $\chi$ is another scalar
field besides inflaton $\varphi$. In our case we may assume that the
symmetry breaking occurs smoothly and the tachyonic mass lasts for a
very short period much smaller than the phase transition time scale
and the Hubble rate. During this period the long wavelength
fluctuations with momentum $k<m$ grow exponentially with $\delta
\chi_{k}$ $\sim \exp(\sqrt{m^2-k^2}~t)$. This situation is known as
{\it tachyonic preheating}.

Our main aim is to estimate the second order metric fluctuations
during these short periods of particle creation occurring in the two
preheating mechanisms.  Enhancement of non-Gaussianity is expected to
take place because of the presence of second order terms sourced by
the first order matter perturbations.  For instant and tachyonic
preheating we need to adopt the multi-field formalism developed in
Refs.~\cite{Antti,we} in order to account for non-Gaussianity due to
the growth of perturbations in $\sigma$ and $\chi$. However as we
shall see in that tachyonic case we can simplify the analysis by
assuming that the inflaton VEV vanishes after inflation, which allows
the metric fluctuations to be mainly seeded by the tachyonic
instability.


\section{Basic equations}

Let us first recapitulate the basic equations required for both
instant and tachyonic preheating. Following Ref.~\cite{Acquaviva}
the metric decomposition is given by,
\begin{eqnarray}
\label{metric}
g_{00} &=& - a(\eta)^2 \left( 1 + 2\phi^{(1)} + \phi^{(2)} \right)\,,\\
g_{0i} &=& 0 \,,\\
g_{ij} &=& a(\eta)^2 \left( 1 - 2\psi^{(1)} - \psi^{(2)} \right) \delta_{ij}\,,
\end{eqnarray}
where the generalized longitudinal gauge is used and the vector and
tensor perturbations are neglected. Here $\eta$ denotes the conformal
time and $a(\eta)$ the scale factor.

In the case of instant preheating we need to study the background
equations of motion for the two fields $\varphi,\sigma$. For
simplicity and for the sake of clarity we assume that the VEV of
$\sigma$ vanishes, $\langle\sigma\rangle=0$, which makes it possible
to obtain analytic approximations from the second order perturbation
equations, as we have shown earlier~\cite{we}.  Such a situation
occurs if $\sigma $ is driven to the minimum of its potential right
after the end of inflation due to a positive Hubble-induced mass
correction, which arises very naturally in many supersymmetric
theories (for a review, see~\cite{Enqvist_rev}).  For our purposes it
is the background motion of $\varphi$ rather than that of $\sigma$
that is important in the instant preheating case. However, it is
interesting to investigate the fluctuations $\delta\sigma$ generated
via the coupling to $\varphi$.

In the tachyonic case for simplicity we will assume an opposite
scenario, where after inflation the inflaton VEV vanishes, $\langle
\varphi\rangle=0$, while the tachyon field $\chi$ is rolling down the
potential. This will allows us to estimate the second order metric
perturbations~\cite{we}. Therefore while describing tachyonic
instability the same equations apply if we replace in the inflaton with the
tachyon in the inflaton equations, $\varphi \rightarrow \chi$, and the second
field with the inflaton in the equations for the second field, $\sigma
\rightarrow \varphi$.

The fields can be divided into the background and perturbation:
\begin{eqnarray}
\label{split}
\varphi &=& \varphi_b(\eta) + \delta^{(1)}\varphi(\eta,{\bf x}) + \frac{1}{2}
\delta^{(2)}\varphi(\eta,{\bf x})\,, \\
\sigma &=& \delta^{(1)}\sigma(\eta,{\bf x}) + \frac{1}{2}
\delta^{(2)}\sigma(\eta,{\bf x}) \,,
\end{eqnarray}
where the background value for $\sigma$ is assumed to vanish.  The
background equations of motion during preheating are then found \cite{Acquaviva, Antti}
to be
\begin{eqnarray}
\label{eqom}
3 {\cal H}^2 &= &\frac{\kappa^2}{2} \varphi_b'^{\,2} +
\frac{1}{2} \kappa^2 a^2 V(\varphi_b) \,,\\
0&=& \varphi_b'' + 2 {\cal H} \varphi_b' + a^2 V^{\prime}(\varphi_b)\,,
\end{eqnarray}
while the $\sigma_b$-equation is trivial. Here $\cal H$ denotes the
Hubble expansion rate expressed in conformal time.

The relevant first order perturbation equations can be written in the
form~\cite{Antti}
\begin{eqnarray}
\phi^{(1)\,''} - \partial_i \partial^i \phi^{(1)} + 2 \left( {\cal H} -
  \frac{\varphi_b''}{\varphi_b'} \right) \phi^{(1)\,'}  + 2
\left( {\cal H}' - \frac{\varphi_b''}{\varphi_b'} {\cal H} \right) \phi^{(1)}
&=& 0 \,,\label{bardeeneq} \\
\delta^{(1)}\sigma'' + 2 {\cal H} \delta^{(1)}\sigma' - \partial_i \partial^i
\delta^{(1)}\sigma + g^2 \varphi_0^2 \, \delta^{(1)}\sigma &=& 0\,.
\label{chieq}
\end{eqnarray}
All the information regarding $\delta^{(1)}\varphi$ is contained in
Eq.~(\ref{bardeeneq}), whose right hand side is zero by virtue of
$\langle \sigma\rangle =0$. Further note that there are no metric
perturbations in Eq.~(\ref{chieq}). This is due to assuming a
vanishing VEV for $\sigma$.  Now the $\sigma$ part can be solved
separately and for the rest the usual single-field results
\cite{Acquaviva} apply.

At the second order we are only interested in the gravitational
perturbation, whose equation can be written in an expanding background
as~\cite{Antti}
\begin{eqnarray}
{\phi^{(2)''}} + 2 \left({\cal H}-\frac{\vp''_b}{\vp'_b}\right)
{\phi^{(2)'}} + 2 \left({\cal H}'- \frac{\vp''_b}{\vp'_b}{\cal H}\right)
{\phi^{(2)}} - \partial_i \partial^i{\phi^{(2)}} = \nonumber \\
{\cal J}_{\varphi,~\textrm{local}} + {\cal J}_{\sigma,~\textrm{local}} 
+ {\cal J}_{\textrm{non-local}}\,, \label{eq:phi2}
\end{eqnarray}
where the source terms $\cal J$ are quadratic combinations of first order
perturbations
\begin{eqnarray}
\label{local0}
{\cal J}_{\varphi, \textrm{local}} &=& \kappa^2 \left[ -2
(\delta^{(1)}\varphi')^2 - 8(\varphi'_b)^2 (\phi^{(1)})^2 + 8
\varphi'_b \phi^{(1)} \delta^{(1)}\varphi' + a^2 \frac{\partial^2
V}{\partial\varphi^2} (\delta^{(1)} \varphi)^2 \right] \nonumber \\ &&
- 24{\cal H}' (\phi^{(1)})^2 - 24 {\cal H} \phi^{(1)}\phi^{(1)\,'}\\
{\cal J}_{\sigma, \textrm{local}} &=& -2 \kappa^2 (\delta^{(1)}
\sigma')^2 + \kappa^2 a^2 \frac{\partial^2 V}{\partial \sigma^2}
(\delta^{(1)} \sigma)^2~\,, \\ 
{\cal J}_{\textrm{non-local}} &=& \triangle^{-1}
f(\delta^{(1)}\varphi,\delta^{(1)}\sigma,\phi^{(1)})\,, \label{non-local}
\end{eqnarray}
where $f$ is a quadratic function of the first order fluctuations and
the coefficients depend on background quantities. Because of the
inverse Laplacian the last source term is non-local.  Typically such
term contains: $\triangle^{-1}(\phi^{(1)\,'}\triangle\phi^{(1)})$,
$\triangle^{-1}(\partial_i\delta^{(1)}\varphi \partial^i
\delta^{(1)}\varphi),\ldots$ Note that the left hand side of
Eq.~(\ref{eq:phi2}) is identical to the first order equation, see
Eq.~(\ref{bardeeneq}).


\section{$f_{NL}^{\phi}$ for Instant Preheating}

Let us now consider a two-field model with the potential
\begin{equation}
\label{maineq}
  V = \frac{1}{2} m_{\varphi}^2 \varphi^2 + g^2 \varphi^2
  \sigma^2\,,
\end{equation}
where $\varphi$ is the inflaton condensate with a mass
$m_{\varphi}$. In instant preheating the particle production occurs
during one oscillation of the inflaton. The particle production occurs
when the inflaton passes through the minimum of the potential
$\varphi=0$. In this case the process can be approximated by writing
\begin{equation}
\label{varphi}
\varphi = \dot\varphi_0 (t-t_0)\,,
\end{equation}
where $\dot\varphi_0$ is the velocity of the field when it passes
through the minimum of the potential at time $t_0$. The time interval
within which the production of $\sigma$ quanta occurs
is~\cite{Lindeinst}
\begin{equation}
\label{time}
\Delta t_* = ( g|\dot\varphi_0|)^{-1/2}\,,
\end{equation}
which is much smaller than the Hubble expansion rate; thus expansion
can be neglected. Note that by virtue of the coupling
$g^2\varphi^2\sigma^2$ the $\sigma$ field acquires a mass and provided that
$g\leq H_{inf}/\varphi\sim 10^{-5}$ for $H_{inf}\sim 10^{13}$~GeV and
$\varphi\sim \kappa^{-1}\sim 10^{18}$~GeV, the fluctuations in
$\sigma$ field were already present on large scales during inflation with
$\delta^{(1)}\sigma\sim H_{inf}/2\pi$.

The occupation number of produced particles jumps from its initial
value zero to a non-zero value during $-\varphi_{\ast}\leq
\varphi\leq \varphi_{\ast}$. In the momentum space the occupation
number is given by~\cite{Lindeinst},
\begin{equation}
\label{occnumber}
n_k = \exp \left( - \frac{\pi k^2}{g |\dot\varphi_0|} \right)\,,
\end{equation}
and the largest number density of produced particles in $x$-space
reads
\begin{equation}
\label{number}
n_{\sigma} \approx \frac{(g|\dot\varphi_0|)^{3/2}}{8\pi^3}\,,
\end{equation}
with the particles having a typical energy of
$(g|\dot\varphi_0|/\pi)^{1/2}$, so that their total energy density is
given by
\begin{equation}
\label{energy}
\rho_{\sigma}\sim \frac{1}{2} (\delta^{(1)}\dot\sigma)^2
 \sim \frac{(g|\dot\varphi_0|)^2}{8\pi^{7/2}} \,.
\end{equation}
These expressions are valid if $m_{\sigma}^2 < g|\dot\varphi_0|$,
a condition that we assume for the rest of our calculation.

Ignoring the expansion of the Universe ($a=1$, $\eta=t$), and using
Eq.~(\ref{varphi}), the second order gravitational perturbation,
Eq.~(\ref{eq:phi2}), is at large scales
\begin{equation}
\label{phi2}
\ddot\phi^{(2)} \sim -2\kappa^2 (\delta^{(1)} \dot\sigma)^2\,.
\end{equation}
The non-Gaussianity in the gravitational potential, parameterized as
$\phi^{(2)} = f_{NL}^{\phi} (\phi^{(1)})^2$, can be estimated by
solving the second order gravitational potential from Eq.~(\ref{phi2})
using Eq.~(\ref{energy}). We obtain
\begin{equation}
\label{nongauss2}
f_{NL}^{\phi} =\left| \frac{g^2|\dot\varphi_0|^2 \Delta t_*^2}
{8 \pi^{7/2} M_P^2 (\phi^{(1)})^2}\right|\,.
\end{equation}
It is a simple exercise to estimate the right hand side for a chaotic
inflaton potential with $m=10^{13}$~GeV and $\phi^{(1)}\sim
10^{-5}$. The velocity of the scalar field at the potential minimum
comes out to be $|\dot\varphi_0|\approx 10^{-7}M_P^2$; using these
values and Eq.~(\ref{time}) we obtain an estimate for the upper limit
of the non-Gaussianity parameter in the case of instant preheating:
\begin{equation}
\label{nongauss3}
f_{NL}^{\phi} \sim 2 g\,.
\end{equation}
Efficient reheating requires $g\geq 10^{-4}$.  However, since
Eq.~(\ref{nongauss3}) implies that $f_{NL}^{\phi}\ll {\cal O}(1)$,
instant preheating is unlikely to yield any detectable non-Gaussian
signal in the forthcoming CMB experiments. The lowest observable value
for $f_{NL}^{\phi}$ by WMAP, Planck and an ideal experiment is
respectively 13.3; 4.7; and 3.5; including polarization data decreases
the limits respectively to 10.9; 2.9; and 1.6 \cite{limits}.


\section{$f_{NL}^{\phi}$ for Tachyonic Preheating}

In order to understand the non-Gaussianity triggered by the tachyonic
instability, let us assume a simple toy model where there is an
inflationary sector and a symmetry breaking phase transition with a
mass squared term as negative:
\begin{equation}
\label{tachyon}
V=V(\varphi)+ V_0 -\frac{1}{2}m^2\chi^2+\frac{\lambda}{4}\,\chi^4 \,.
\end{equation}
We assume that the inflaton potential is some polynomial potential
with a vanishing VEV, $V(\varphi)\sim f(\varphi^{n})$. Inflation is
supported by both $V(\varphi)+V_0$. During inflation we assume that the
tachyon field is sitting at the maximum $\chi=0$ by virtue of large
friction. The mass of $\chi$ is such that the tachyonic instability is
triggered when $m\geq H\sim V_{0}/3M_{P}^2$. During this period we
assume that the inflaton settles down to $\langle \varphi\rangle =0$. This
will allow us to separate the tachyon fluctuations from that of the
inflaton. This also allows us to use the same equations
(\ref{split} - \ref{non-local}) but now the tachyon field $\chi$ obeys
Eq.~(\ref{bardeeneq}) and the inflaton field $\varphi$ obeys
Eq.~(\ref{chieq}). So we replace $\varphi\rightarrow \chi$ in
Eq.~(\ref{bardeeneq}) and in $\sigma\rightarrow \varphi$ in Eq.~(\ref{chieq}),
and we make similar replacements in Eq.~(\ref{eq:phi2}) and the expressions
for ${\cal J}_{\chi, \textrm{local}}$ and ${\cal J}_{\varphi, \textrm{local}}$
respectively.

The rolling of a tachyon in itself results in an exponential
instability in the perturbations of $\chi$ with physical momenta
smaller than the mass. The tachyonic growth takes place within a short
time interval, $t_{\ast}\sim (1/2m)\ln(\pi^2/\lambda)$
(see~\cite{Lindetach}).  During this short period the occupation
number of $\chi$ quanta grows exponentially for modes $k<m$ up to
$n_{k}\sim \exp(2mt_{\ast})\sim \exp(\ln(\pi^2/\lambda))\sim
\pi^2/\lambda$.  For very small self-coupling, which is required for a
successful inflation, the occupation number, which depends inversely
on the coupling constant, can become much larger than one.

The scalar field fluctuations, which are responsible for exponentially
enhancing the occupation number for $\chi$ quanta, also couple to the
metric fluctuations. If we assume that the modes grow within a time
interval much smaller than the Hubble rate, we can set ${\cal H} = 0$
in Eq.~(\ref{bardeeneq}). Then, in the long wavelength limit, we get
from Eq.~(\ref{bardeeneq}),
\begin{equation}
\label{bardeeneq2}
\ddot\phi^{(1)} - 2A\, \dot\phi^{(1)} = 0\,,
\end{equation}
where $A=\ddot\chi_0/\dot\chi_0$. With the assumption of a brief
tachyonic stage, $\dot\chi_0,\,\ddot\chi_0$ are effectively
constants. Note that although during rolling tachyon the long
wavelength modes are excited, but it is important that the tachyon
perturbations must exist during inflation. In this respect the tachyon
fluctuations are isocurvature in nature. In order to further simplify
our calculation we neglect the inflaton perturbations in our
subsequent analysis.

There are two solutions of Eq.~(\ref{bardeeneq2}); a constant
$\phi^{(1)} \sim 10^{-5}$, and an exponentially growing solution
$\phi^{(1)} \propto \exp(2A t)$. The former case arises when we
recognize the constant by the temperature anisotropy of the observed
CMB fluctuations. If the isocurvature component at the end of
inflation is small, then the first derivative of $\phi^{(1)}$ is also
small but non-vanishing. Hence we may neglect the exponential solution
of the first order metric perturbation. Although our argument holds
good for the first order metric perturbations, but as we shall show
this will not be the case for the second order calculation.  With
these simplified approximations we can then estimate the amount of
generated non-Gaussianity by following a logic similar to the case of
instant preheating.

First, the number density of the produced particles in $x$-space is
given by $n_\chi\sim m^3/(8\pi\lambda)$. Hence the total energy
density stored in produced $\chi$ quanta is given by
\begin{equation}
\label{energy2}
\rho_{\chi}\sim \frac{1}{2} (\delta^{(1)}\dot\chi)^2
 \sim m n_{\chi} \sim \frac{1}{8 \pi} \frac{m^4}{\lambda}\,.
\end{equation}
The equation for the second order metric perturbation now includes a
source term which includes ${\cal J}_{I, \textrm{local}}$ and ${\cal
J}_{\chi, \textrm{local}}$. However the main contribution comes from
the excitations of the tachyonic instability. The inflaton
fluctuations are subdominant compared to the exponential growth of
$\delta^{(1)}\chi$, when it is settled around its VEV $\langle
\varphi\rangle=0$. In the the long wavelength regime the perturbation
equation reads as
\begin{equation}
\label{phi22} \ddot\phi^{(2)} \sim -
\frac{1}{\pi}\kappa^2 \frac{m^4}{\lambda}\,.
\end{equation}
Integrating the above equation over the time interval
$t_{\ast}\sim(1/2m)\ln(\pi^2/\lambda)$, we find $\phi^{(2)}\sim
(m/M_P)^2\ln^2(\pi^2/\lambda)/(4\pi\lambda)$. The non-Gaussianity
parameter for tachyonic preheating in case the first order metric
perturbation stays constant is then roughly given by
\begin{equation}
\label{fnltach}
f_{NL}^{\phi}\sim\frac{1}{4\pi}\Big(\frac{m}{M_P}\Big)^2\frac{1}{\lambda}\frac{1}{(\phi^{(1)})^2}\ln^2\left(\frac{\pi^2}{\lambda}\right)\,,
\end{equation}
where we substitute $\kappa \sim M_{P}^{-1}$. Writing this in terms of
$V_0={m^4}/(4\lambda)$ and taking $\phi^{(1)}\sim 10^{-5}$, we obtain
\begin{equation}
\label{obslim}
f_{NL}^{\phi}\sim 1.6\times
10^9\;\lambda^{-1/2}\;\left(\frac{V_0^{1/4}}{M_P}\right)^2
\ln^2\left(\frac{\pi^2}{\lambda}\right)\,.
\end{equation}
This expression should be compared with
the observationally constrained one: $f_{NL}=-f^{\phi}_{NL}+11/6$, see
for instance \cite{ngrev}.\footnote{The measure of non-Gaussianity is
the non-linearity parameter $\phi^{(2)}=f_{NL}^{\phi}
(\phi^{(1)})^2$. In general $f_{NL}^{\phi}$ contains momentum
dependent part, i.e. $f_{NL}^{\phi}({\bf k_1},{\bf k_2})$, and the
constant piece. It is the non-local terms which affect the momentum
dependent part, since all the derivatives are replaced by momenta in
the Fourier space. However the present constraint on non-Gaussianity
parameter from WMAP does not give the momentum dependent constraint
but only the constant part. Therefore the non-local terms do not lead
to any observable constraints, so we do not consider them here.}
The current observational limit of the non-linearity parameter set by
WMAP is $-132 <f_{NL}^{\phi}<60$, at $95\%$ confidence
level~\cite{WMAP-nongaussian}.
Adopting the upper limit $|f_{NL}^\phi|<132$ and rearranging, we
arrive at the bound ${V_0^{1/4}/ M_P}\lsim 3\times
10^{-4}\lambda^{1/4}\ln^{-1}\left({\pi^2} /{\lambda}\right)\,$.

For an effective field theory to remain perturbative we should require
that $\lambda\ll 1$, which yields the interesting constraint
$V_{0}^{1/4}/M_{P}\ll 10^{-4}$. Note that compared to the usual bound
$V_{0}^{1/4} \leq 10^{16}$~GeV from COBE normalization, the absence of
observable non-Gaussianity implies a bound on the scale of tachyonic
instability $V_{0}^{1/4}$ which is more stringent by two orders of
magnitude.  The parameter space allowed by WMAP data is given by the
region below the red curve in Figure~\ref{tachyonfig}.


\begin{figure}
\begin{center}
\includegraphics[width=0.75\textwidth]{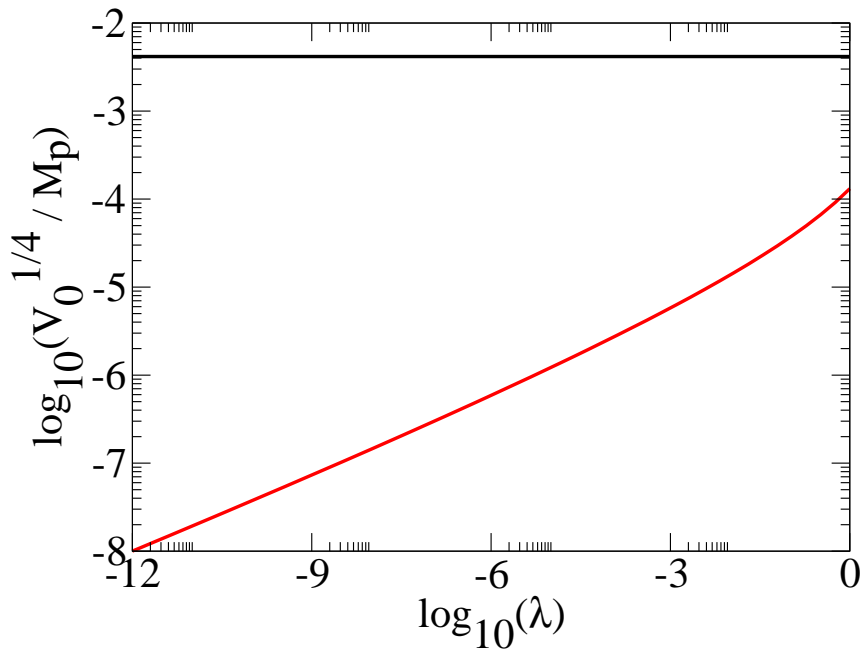}
\caption{\small The parameter space in tachyonic preheating
allowed by the WMAP data
for single field inflation (below the red
curve). Black horizontal line is the usual limit $V_0^{1/4} = 10^{16}$ GeV
coming from the COBE normalization.}\label{tachyonfig}
\end{center}
\end{figure}


When obtaining the result above we have assumed that the first order
metric perturbations is roughly given by the constant value determined
by the inflationary epoch. Let us investigate the other limit when the
first order metric fluctuations also obtain an exponentially growing
solution by virtue of the tachyon excitations. To check this
possibility, let us assume that the exponential solution actually
dominates over the constant one. Following the second order analysis
and assuming that the main contribution to the second order
perturbation arises from the tachyonic instability, we obtain from
Eq.~(\ref{eq:phi2})
\begin{eqnarray}
\label{bardeeneq3}
 \ddot\phi^{(2)} - 2A\,\dot\phi^{(2)} = -\kappa^2
\left[ 2(\delta^{(1)}\dot\chi)^2
+ 8\dot\chi_0^2 (\phi^{(1)})^2 - V_{,\chi\chi}
(\delta^{(1)} \chi)^2 - 8 \dot\chi_0 \phi^{(1)}
\delta^{(1)}\dot\chi \right],
\end{eqnarray}
where $V_{,\chi\chi} = -m^2 + 3\lambda\chi^2$ and where
$\delta^{(1)}\chi$ can be solved through the Einstein
constraint~\cite{ngrev} $\delta^{(1)}\chi = \frac{2}{\kappa^2
\dot\chi_0} \left(\dot\phi^{(1)} + H \phi^{(1)} \right)$.

For the tachyonic region $3\lambda\chi^2<m^2$ so that we can take
$V_{,\chi\chi}\sim -m^2$. Now $\phi^{(2)}$ contains the homogeneous
solution $\sim \exp(2At)$ together with a source part $\sim
\exp(4At)$~\footnote{The second order metric perturbations always have
a growing source term by virtue of the non-vanishing background motion
of the scalar field, i.e. $\delta^{(1)}\chi$,
$\delta^{(1)}\dot\chi$.}.  After a while the source part dominates and
we obtain the result\footnote{Assuming $A$ is constant; in reality
there will be a small time variation but we may assume that most of
the interesting modes are growing within a time interval which is
short compared to the variation in $\ddot\chi_{0}/\dot\chi_{0}$ and
the Hubble rate.}
\begin{equation}
\label{param}
f_{NL}^{\phi}= \frac{\phi^{(2)}}{(\phi^{(1)})^2} =\left|8 - \frac{2m^2}
{\kappa^2\dot\chi_0^2} - \frac{4 \ddot\chi_0^2}{
\kappa^2\dot\chi_0^4} - \frac{\kappa^2 \dot\chi_0^4}
{\ddot\chi_0^2}\right |\,.
\end{equation}
The tachyonic growth persists until $\chi\sim m/(2\sqrt{\lambda})$,
for a time span $t_{\ast} \sim (1/2m) \ln(\pi^2/\lambda)$
~\cite{Lindetach}. With these approximations, writing
$V_{0}=m^4/(4\lambda)$, we obtain
\begin{equation}
\label{param2}
f_{NL}^{\phi} \approx 8 - 8 \sqrt{\frac{\lambda M_P^4}{V_0}} -
 \frac{1}{2} \sqrt{\frac{V_0}{\lambda M_P^4}} - \sqrt{\frac{\lambda
 M_P^4}{V_0}} {\rm ln}^2 \left( \frac{\pi^2}{\lambda} \right) \,.
\end{equation}
If we assume COBE normalization $V_0^{1/4}\leq 10^{16}$~GeV, the
minimum of $V_0$ is given by the conditions ${V_0^{1/4}}/{M_P}=
\lambda^{1/4} \sqrt{8-f_{NL}^{\phi}}$ and $\lambda = \pi^2 \exp ( -
\sqrt{(8-f_{NL}^{\phi})^2/2 - 8} )$.  These imply the limit
$f_{NL}^{\phi}<-37$ regardless of the value of $\lambda$, well within
the observational capabilities of WMAP.


Many string-motivated inflationary models could thus be constrained by
the present and future limits on non-Gaussianity.  An example would be
inflation first driven by brane-anti-brane interaction and then coming
to an end when the tachyonic instability is triggered~\cite{Cliff}.
Should we take the tachyon coupling to be very small $\lambda\sim
10^{-12}$, as constrained by the amplitude of the CMB scalar
fluctuations~\cite{WMAPinflation}, we immediately obtain from
Eq.~(\ref{obslim}) that $V_{0}^{1/4}/M_{P}\leq 10^{-7}$ in order to
comply with the current WMAP limit on non-Gaussianity.  Such
considerations imply very interesting constraints on the scale of the
tachyonic instability and on the tachyon self coupling in
brane-anti-brane driven inflation.

In obtaining our limits we have made some approximations. In
particular, we assumed that the VEV of $\sigma$ field is vanishing in
the instant preheating case and in tachyonic preheating we assumed
that the inflaton VEV is vanishing after the end of inflation. These
assumptions make our analysis simple and provide a handle on the
non-Gaussianity parameter.


\vskip40pt

We are thankful to Robert Brandenberger, Cliff Burgess, Andrew Liddle,
David Lyth, Licia Verde and Filippo Vernizzi for discussions.  A.V.~is
supported by the Magnus Ehrnrooth Foundation. A.V.~thanks NORDITA and
NBI for their kind hospitality during the course of this work. K.E. is
supported in part by the Academy of Finland grant no. 75065.

\vskip 30pt


\begin{thebibliography}{99}

\bibitem{Robert}
J.~H.~Traschen and R.~H.~Brandenberger,
Phys.\ Rev.\ D {\bf 42}, 2491 (1990).

\bibitem{Many1}
L.~Kofman, A.~D.~Linde and A.~A.~Starobinsky,
Phys.\ Rev.\ Lett.\  {\bf 76}, 1011 (1996)
[arXiv:hep-th/9510119];
Y.~Shtanov, J.~H.~Traschen and R.~H.~Brandenberger,
Phys.\ Rev.\ D {\bf 51}, 5438 (1995)
[arXiv:hep-ph/9407247];
D.~Boyanovsky, H.~J.~de Vega and R.~Holman,
arXiv:hep-ph/9701304, and refs. therein.

\bibitem{Linde}
L.~Kofman, A.~D.~Linde and A.~A.~Starobinsky,
Phys.\ Rev.\ D {\bf 56}, 3258 (1997)
[arXiv:hep-ph/9704452].

\bibitem{Dan}
D.~Cormier, K.~Heitmann and A.~Mazumdar,
Phys.\ Rev.\ D {\bf 65}, 083521 (2002)
[arXiv:hep-ph/0105236].

\bibitem{Jaikumar}
P.~Jaikumar and A.~Mazumdar,
Nucl.\ Phys.\ B {\bf 683}, 264 (2004)
[arXiv:hep-ph/0212265];
K.~Enqvist and J.~H\"ogdahl,
arXiv:hep-ph/0405299.



\bibitem{Many2}
B.~A.~Bassett, D.~I.~Kaiser and R.~Maartens,
Phys.\ Lett.\ B {\bf 455}, 84 (1999)
[arXiv:hep-ph/9808404];
B.~A.~Bassett, F.~Tamburini, D.~I.~Kaiser and R.~Maartens,
Nucl.\ Phys.\ B {\bf 561}, 188 (1999)
[arXiv:hep-ph/9901319];
K.~Jedamzik and G.~Sigl,
Phys.\ Rev.\ D {\bf 61}, 023519 (2000)
[arXiv:hep-ph/9906287];
F.~Finelli and R.~H.~Brandenberger,
Phys.\ Rev.\ Lett.\  {\bf 82}, 1362 (1999)
[arXiv:hep-ph/9809490];
F.~Finelli and R.~H.~Brandenberger,
Phys.\ Rev.\ D {\bf 62}, 083502 (2000)
[arXiv:hep-ph/0003172];
A.~R.~Liddle, D.~H.~Lyth, K.~A.~Malik and D.~Wands,
Phys.\ Rev.\ D {\bf 61}, 103509 (2000)
[arXiv:hep-ph/9912473].

\bibitem{Easther}
M.~Parry and R.~Easther,
Phys.\ Rev.\ D {\bf 59}, 061301 (1999)
[arXiv:hep-ph/9809574];
R.~Easther and M.~Parry,
Phys.\ Rev.\ D {\bf 62}, 103503 (2000)
[arXiv:hep-ph/9910441];
F.~Finelli and S.~Khlebnikov,
Phys.\ Lett.\ B {\bf 504} (2001) 309
[arXiv:hep-ph/0009093];
F.~Finelli and S.~Khlebnikov,
Phys.\ Rev.\ D {\bf 65} (2002) 043505 [arXiv:hep-ph/0107143].

\bibitem{MFB}
V.~F.~Mukhanov, H.~A.~Feldman and R.~H.~Brandenberger,
Phys.\ Rept.\  {\bf 215}, 203 (1992).




\bibitem{ngrev}
N.~Bartolo, E.~Komatsu, S.~Matarrese and A.~Riotto,
Phys.\ Rept.\  {\bf 402} (2004) 103
[arXiv:astro-ph/0406398].

\bibitem{we}
K.~Enqvist, A.~Jokinen, A.~Mazumdar, T.~Multam\"aki and A.~V\"aihk\"onen,
arXiv:astro-ph/0411394.



\bibitem{Acquaviva}
V.~Acquaviva, N.~Bartolo, S.~Matarrese and A.~Riotto,
Nucl.\ Phys.\ B {\bf 667} (2003) 119
[arXiv:astro-ph/0209156].

\bibitem{Antti}
K.~Enqvist and A.~V\"aihk\"onen,
JCAP {\bf 0409}, 006 (2004)
[arXiv:hep-ph/0405103].




\bibitem{Enqvist}
K.~Enqvist, S.~Kasuya and A.~Mazumdar,
Phys.\ Rev.\ Lett.\  {\bf 90}, 091302 (2003)
[arXiv:hep-ph/0211147];
K.~Enqvist, A.~Jokinen, S.~Kasuya and A.~Mazumdar,
Phys.\ Rev.\ D {\bf 68}, 103507 (2003)
[arXiv:hep-ph/0303165];
K.~Enqvist, S.~Kasuya and A.~Mazumdar,
Phys.\ Rev.\ Lett.\  {\bf 93}, 061301 (2004)
[arXiv:hep-ph/0311224];
K.~Enqvist, A.~Mazumdar and A.~Perez-Lorenzana,
arXiv:hep-th/0403044.



\bibitem{Riotto}
N.~Bartolo, S.~Matarrese and A.~Riotto,
JHEP {\bf 0404}, 006 (2004)
[arXiv:astro-ph/0308088].

\bibitem{Lindeinst}
G.~N.~Felder, L.~Kofman and A.~D.~Linde,
Phys.\ Rev.\ D {\bf 59}, 123523 (1999)
[arXiv:hep-ph/9812289].


\bibitem{Lindetach}
G.~N.~Felder, J.~Garcia-Bellido, P.~B.~Greene, L.~Kofman, A.~D.~Linde and I.~Tkachev,
Phys.\ Rev.\ Lett.\  {\bf 87}, 011601 (2001)
[arXiv:hep-ph/0012142].


\bibitem{Enqvist_rev}
K.~Enqvist and A.~Mazumdar,
Phys.\ Rept.\  {\bf 380}, 99 (2003)
[arXiv:hep-ph/0209244].


\bibitem{limits}
D.~Babich and M.~Zaldarriaga,
Phys.\ Rev.\ D {\bf 70} (2004) 083005
[arXiv:astro-ph/0408455].


\bibitem{WMAP-nongaussian}
E.~Komatsu {\it et al.},
Astrophys.\ J.\ Suppl.\  {\bf 148} (2003) 119
[arXiv:astro-ph/0302223].

\bibitem{Cliff}
C.~P.~Burgess, M.~Majumdar, D.~Nolte, F.~Quevedo, G.~Rajesh and R.~J.~Zhang,
JHEP {\bf 0107}, 047 (2001)
[arXiv:hep-th/0105204];
A.~Mazumdar, S.~Panda and A.~Perez-Lorenzana,
Nucl.\ Phys.\ B {\bf 614}, 101 (2001)
[arXiv:hep-ph/0107058].




\bibitem{WMAPinflation}
H.~V.~Peiris {\it et al.},
Astrophys.\ J.\ Suppl.\  {\bf 148} (2003) 213
[arXiv:astro-ph/0302225].



\end{thebibliography}
\end{document}